# Title

Lithium depth profiling in NMC/Graphite commercial coin cells under high C-rate cycling

# Authors

Naisargi Kanabar[1] ( nkanabar@albany.edu ) - Corresponding Author

Seiichiro Higashiya[2] ( shigashiya@albany.edu )

Daniele Cherniak[1] ( dcherniak@albany.edu )

Devendra Sadana[2] ( dsadana@albany.edu )

Stephen Bedell[2] ( sbedell@albany.edu )

Haralabos Efstathiadis[2] ( hefstathiadis@albany.edu )

# Affiliations

[1]Department of Physics, University at Albany (SUNY), Albany, NY 12222

[2]Department of Nanoscale Science and Engineering, University at Albany (SUNY), Albany, NY 12222

# Abstract

This study examines the distribution and evolution of lithium in both anode and cathode materials of commercial lithium-ion coin cells subjected to high C-rate cycling, providing insights into the mechanisms of lithium loss, trapping, and plating. Cells were cycled at 1C to 3C rates, and post-mortem analysis was performed using Li nuclear reaction analysis (Li-NRA), x-ray diffraction (XRD), and scanning electron microscopy (SEM) equipped with energy-dispersive x-ray spectroscopy (EDS). Li-NRA, using the resonant nuclear reaction between an incident high-energy proton and lithium, was used to measure the depth distribution of Li in the cathode and anode layers. The Li-NRA analysis revealed a surface lithium peak on the anode, likely associated with SEI formation and lithium plating, while the cathode exhibited a decrease in lithium content by ~19.7%. XRD analysis of the cycled cathode showed an expansion of the c-lattice parameter and peak shifts consistent with lithium depletion and structural deformation, supported by SEM imaging. In contrast, the dead graphite anode shows an enhanced peak at 43.3°, which corresponds to the presence of $Li_2Co_3$. 3-C rate cycling also led to capacity fade and an increase in internal resistance, highlighting the impact of lithium plating on cell performance.

# Graphical Abstract

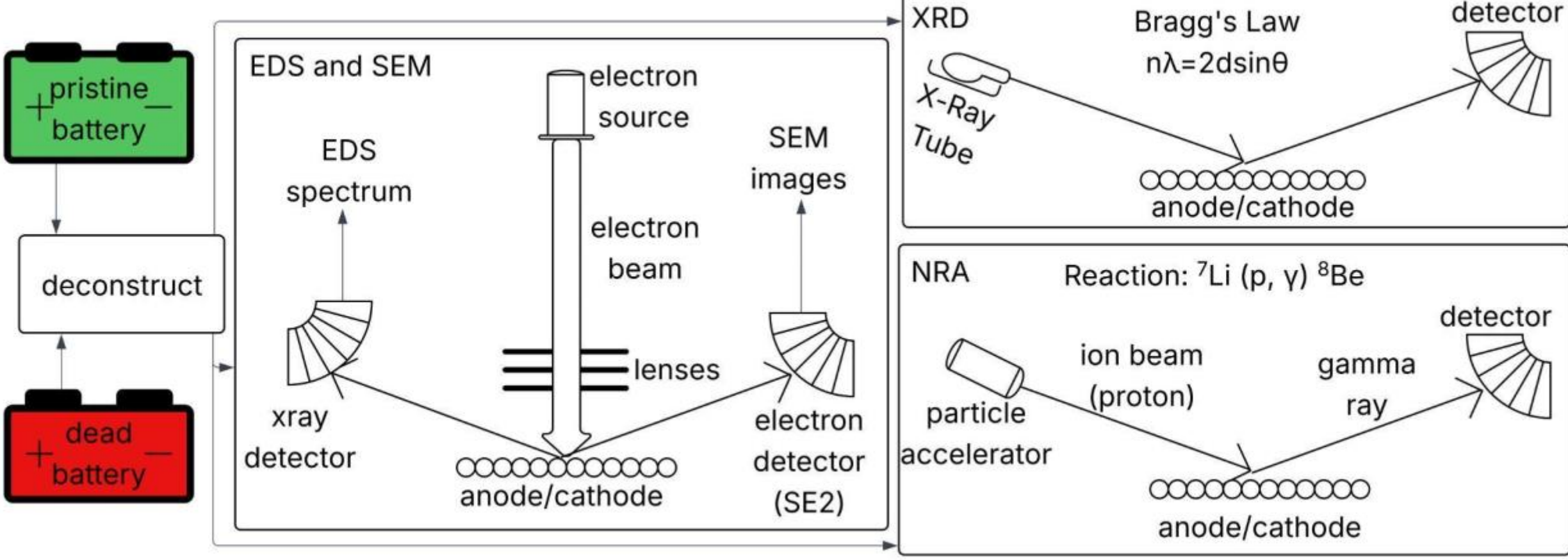

# 1. Introduction

Rechargeable lithium-ion batteries (LIBs) are becoming increasingly vital to modern society by powering portable electronics and electric vehicles, while also aiding the transition to a low-carbon economy. However, longer charging time is slowing down the adoption of electric vehicles (EVs). At present, it takes about 30 minutes to recharge a battery to 80% of its full capacity [1]. To meet consumer demand, one approach is to use a high C-rate. The C-rate relates the charging current to the capacity of the active material, defining the time it should take for the cell to fully charge or discharge. However, performance under high C-rate cycling remains a critical limitation, as high C-rate operation can induce complex degradation mechanisms including lithium inventory loss, electrode structure destabilization, lithium plating, and accelerated solid electrolyte interphase (SEI) growth, all of which contribute to capacity fade and impedance increase [2–4].

To enhance the energy density of LIBs for EVs, manufacturers have progressively increased the nickel content in $LiNi_{1-x-y}Mn_xCo_yO_2$ (NMC) cathode materials. This compositional evolution has advanced from NMC111 ($LiNi_{0.33}Mn_{0.33}Co_{0.33}O_2$) to NMC532 ($LiNi_{0.50}Mn_{0.30}Co_{0.20}O_2$), NMC622 ($LiNi_{0.60}Mn_{0.20}Co_{0.20}O_2$), and most recently NMC811 ($LiNi_{0.80}Mn_{0.10}Co_{0.10}O_2$). Despite the higher capacity provided by Ni-rich NMC materials, they are prone to structural degradation caused by lattice instability, oxygen release at high states of charge, and increased susceptibility to microcracking, all of which negatively impact long-term cycling stability and overall battery lifespan [5-8]. For graphite anodes, lithium plating and SEI thickening are key concerns during high-rate cycling, especially at low temperatures or high states of charge [SOC] [9-10].

Understanding how lithium-ion batteries degrade under high c-rates requires accurate techniques to study electrode composition. Methods such as time-of-flight secondary ion mass spectrometry (TOF-SIMS) [11], X-ray photoelectron spectroscopy (XPS) [12], and inductively coupled plasma mass spectrometry (ICP-MS) [13] are commonly used, but they are destructive because they rely on sputtering, and they often cannot measure lithium well due to its low atomic number. Nuclear Reaction Analysis (NRA), on the other hand, is a mostly non-destructive method that can directly detect mobile lithium in electrodes with high accuracy.

Previous studies have demonstrated the utility of the Li-NRA in Li depth profiling. Sunitha et al. [14] used the $^{7}Li(p,\gamma)^{8}Be$ nuclear reaction to profile Li in $LiCoO_2$ and graphite, achieving accuracy within 0.2 at.%. Schulz et al. [15] applied the same approach to NMC (442) and graphite, while other cathode materials, such as $LiNiVO_4$, $LiFeO_2$, and $MoO_3$, have also been studied for their Li contents [16-18]. However, to the best of the authors' knowledge, Li-NRA has not previously been applied to commercial NMC/graphite full cells subjected to high C-rate cycling until the discharge capacity declined to 20% of its initial value. This enables direct assessment of lithium loss mechanisms under rate-driven degradation conditions that are highly relevant to practical, commercial battery operation.

In this study, we examined commercial NMC/graphite coin cells (3032) cycled at 1C - 3C rates. A combination of techniques was used: x-ray diffraction (XRD) to track lattice parameter shifts, phase changes, and peak broadening related to lithium intercalation/deintercalation [19]; Scanning electron microscopy (SEM) to observe surface features such as cracking, roughness, and SEI buildup [20]; Energy-dispersive x-ray spectroscopy (EDS) to measure electrode density and composition, which are used for stopping and range of ions in matter (SRIM) calculations; and Li-NRA to directly measure lithium in both pristine and cycled electrodes [21].

The formation and growth of the SEI reduces the total available lithium for the battery operation, resulting in capacity loss, increased internal resistance, and reduced power density [22]. According to Li et al. [23], charge rates exceeding 2C promote lithium plating in the anode region and increase diffusion-induced stress in graphite particles, resulting in loss of active material being the main driver of accelerated capacity fade. Using Li-NRA, we can quantify how much the lithium concentration reduces at the surface and in the bulk of the cathode material. Most importantly, Li-NRA is a highly sensitive method for quantitative lithium profiling with depth resolution, which is critical for determining lithium retention and distribution across the electrode thickness.[24]

Battery cycling at high charge rates imposes considerable mechanical stress on both cathode and anode materials, driven by the accelerated intercalation and deintercalation of lithium ions. These rapid ionic movements induce repeated volumetric changes in the electrodes, which can ultimately lead to micro-cracking of the crystal structure and macro-scale fractures throughout the electrode. [25,26]

By integrating structural analysis through XRD, morphological assessment via SEM, and quantitative lithium profiling using Li-NRA, this study delivers a comprehensive evaluation of the degradation processes in NMC and graphite electrodes subjected to high C-rate cycling. The combined insights enable a clearer understanding of lithium loss mechanisms and their relationship to structural and surface-level damage. Ultimately, these findings contribute to advancing battery design strategies and operational guidelines aimed at enhancing the durability and performance of lithium-ion batteries under high-stress cycling conditions.

# 2. Experimental

## 2.1. Cell Type

In this study, high-energy commercial lithium-ion coin cells with a nominal capacity of 200mAh (3032) were utilized. The full cell consisted of nickel-rich Manganese Cobalt Oxide as the cathode and graphite as the anode. The composition information is provided by the supplier and is listed in Table S5.

## 2.2. Electrochemical cycling test

Electrochemical measurements were conducted using the battery tester Gamry Reference 600+. NMC/Graphite coin cells (1C = 200 mA) were charged at 1C, 2C, and 3C, eventually failing after extended cycles at high C-rates. For extended cycling, the commercial coin cells were charged using a constant current (CC) up to 4.3 V and then discharged at the same C-rates within a voltage range of 4.3–2.7 V. All tests were carried out at room temperature (25°C).

After cycling, the coin cells were disassembled in a discharged state for safety inside a glovebox (0.5 ppm $H_2O$ and 0.2 ppm $O_2$). The electrode material was extracted from the same location for all samples. The extracted material was then sealed in an airtight pouch inside the glovebox before proceeding to the characterization tools.

## 2.3. Characterization of electrodes

X-ray diffraction (XRD) data were collected using a Rigaku SmartLab SE multipurpose θ-θ X-ray diffractometer with Cu Kα radiation and a HyPix-400 two-dimensional advanced photon counting hybrid pixel array detector at 20 °C. The unit-cell parameters were calculated using a Le Bail fit of the powder XRD patterns performed in SmartLab Studio II software (version 4.6.411.0).

The morphology, microstructure, and composition of the samples were examined using a Zeiss Leo 1550 scanning electron microscope (SEM) equipped with a backscatter detector and energy-dispersive X-ray spectroscopy (EDS), operating at an accelerating voltage of 20 kV.

## 2.4. Lithium nuclear reaction analysis

Lithium Nuclear Reaction Analysis (Li-NRA) of both the negative and positive electrodes was conducted using the nuclear reaction $^{7}Li(p, \gamma)^{8}Be$ with a resonance energy of 0.44 MeV. The incident beam energy was varied between 0.44 MeV and 0.70 MeV in 10 keV increments, and the resulting γ-ray intensity was collected using a bismuth germanium oxide (BGO) scintillation detector positioned on the 30° beamline until a beam charge of 2 μC was accumulated. A bias voltage of 0.80 kV was applied to the BGO detector to adjust the peak location of Li, and a beam current of 30 to 60 nA was maintained throughout the experiment. Depth profiling was performed at the Ion Beam Lab at SUNY Albany. More information about the facility and technique can be found in Lanford et al. [27,28].

## 2.5. Calculation of concentration and depth

The intensity of the characteristic gamma-ray signal (ROI integral) is directly proportional to the lithium content, while a stepwise increase in proton energy enables depth profiling of the sample. To obtain Li at% versus depth, we first convert the raw data of γ-ray counts versus energy. By calculating the SRIM table for both the standard sample and the electrode material, we obtain the nuclear and electronic stopping powers (dE/dx) for each energy. These values allow us to determine the depth x corresponding to each incident beam energy E, using the equation $x = \frac{E-E_{res}}{dE/dx}$, where E is the incident beam energy, $E_{res}$ is the resonance energy, and (dE/dx) is the stopping power of the element of interest in the target. To calculate the Li concentration, we use Equation $Li\ Concentration(x) = \left(\frac{(\frac{dE}{dx})_{std}}{(\frac{dE}{dx})_{sample}}\right)\left(\frac{ROI_{sample}}{ROI_{std}}\right)\rho_{std}^{Li}$, where ROI is the γ-ray counts and $\rho^{Li}_{std}$ is the Li concentration in the standard sample. Here, we used $LiTaO_3$ as the standard sample.

# 3. Results and Discussion

## 3.1. Cycling of the full cells

In this study, a total of 278 charge-discharge cycles were performed between 2.7 V and 4.3 V at room temperature (25 °C). Initially, the cell was cycled at a current of 200 mA (1C charge/1C discharge) for 11 cycles, during which the cell capacity slightly exceeded 200 mAh. Subsequently, the current was increased to 400 mA (2C charge/2C discharge) for 67 cycles, followed by a further increase to 600 mA (3C charge/3C discharge) for the remaining 200 cycles. Under the high c-rate (3C) cycling conditions, the cell capacity progressively declined, reaching approximately 42 mAh by the end of the test. (Fig. 1a)

Jia Guo et al. [29] reported an early-stage capacity increase in commercial NMC/graphite Li-ion batteries, particularly at large depths of discharge (75–100%). This behavior was attributed to enhanced $Li^+$ insertion and improved diffusion kinetics in the graphite anode, driven by structural evolution such as increased interlayer spacing and partial nanosheet exfoliation. These changes reduce $Li^+$ insertion resistance and expand the accessible electrochemical window, leading to a temporary capacity gain. This mechanism is consistent with the slightly higher than 200 mAh capacity observed during the early cycling stage in this study.

At a 1C rate, the cell exhibited an initial charge capacity of 206.50 mAh and retained a reversible capacity of 207.06 mAh after 11 cycles, indicating a slight activation-related capacity increase in the early stage. Capacity retention within this 1C block, excluding the first transition cycle, showed negligible fade with approximately 100.3% retention. When the cycling rate increased to 2C, the initial charge capacity decreased to 148.97 mAh, and the cell retained a reversible capacity of 132.84 mAh after 67 cycles, corresponding to an average capacity fade of 0.24 mAh per cycle (0.162% per cycle) and 89.2% retention. Further increasing the rate to 3C resulted in an initial charge capacity of 82.16 mAh, with the cell retaining a reversible capacity of 42.67 mAh after 200 cycles. This corresponds to an average fade of 0.19 mAh per cycle (0.240% per cycle) and a retention of 51.9%. These results demonstrate that higher C-rates accelerate the initial capacity drop and lead to poorer capacity retention, even when the absolute per-cycle loss in mAh may appear comparable. This is attributed to the smaller accessible capacity at higher rates and the greater relative fade observed.

Therefore, a comprehensive understanding of lithium distribution at both the surface and within the bulk of the electrodes is essential. The Li-NRA technique enables quantitative insight into the spatial distribution of lithium in the positive and negative electrodes. Detailed analysis of the lithium depth profiles obtained from Li-NRA measurements is presented in Section 3.3.

In this work, the first-cycle coulombic efficiency (CE) after each C-rate change was excluded from analysis due to transient artifacts that produced abnormally high values. These anomalies arose from a mismatch between the cell's state of charge at the end of the previous rate and the initial conditions of the new rate, leading to a lower measured charge capacity relative to the subsequent discharge capacity. Such effects are generally attributed to electrode relaxation, lithium redistribution between electrodes, and kinetic limitations during rate transitions in NMC/graphite cells. Since these first-cycle CE values do not represent the steady-state electrochemical efficiency, only stabilized CE values from subsequent cycles at each C-rate are reported and discussed.

Fig. 1b exhibits the CE of the NMC/graphite coin cell over cycling. During cycling at 1C, the CE remained highly stable, averaging 99.88%, reflecting minimal parasitic reactions and efficient lithium shuttling between electrodes. At higher C-rates, a slight decrease in the average CE was observed (99.83% at 2C & followed by 99.80% at 3C), which can be attributed to increased polarization and possible electrolyte decomposition under higher current densities. The overall stability of CE across different cycling conditions correlates with the observed capacity retention trends, suggesting that the cell degradation is primarily driven by gradual loss of active lithium inventory rather than continuous parasitic reactions.

The internal resistance (IR) evolution of the NMC/graphite coin cell was evaluated across different C-rate regimes, measured at the transition between the end of charge and the start of discharge (Fig. 2). At 1C, the average internal resistance stabilized at approximately 0.32 Ω. Increasing the rate to 2C yielded a slightly lower initial resistance of 0.29 Ω; notably, during the second cycle at this rate, the resistance further decreased to 0.22 Ω. This reduction is likely associated with the increased discharge current, which leads to greater joule heating ($\propto I^2R$) and a raise in cell temperature during measurement [30]. Elevated temperature enhances electrolyte ionic conductivity and lithium-ion transport, thereby improving charge-transfer kinetics and reducing the ionic and polarization components of the IR. These thermal effects at higher C-rates thus provide a plausible explanation for the observed decrease in IR. Upon switching to 3C, the internal resistance initially measured 0.23 Ω but exhibited an approximately linear increase over the subsequent 200 cycles, ultimately reaching ~0.34 Ω. This progressive rise in resistance at 3C indicates that accelerated interfacial aging, increased polarization, and transport limitations outweigh the beneficial thermal effects. This aligns with the observed decline in reversible capacity, indicating cumulative degradation of electrode-electrolyte interfaces and possible loss of electrical contact within the electrode microstructure.

## 3.2. Structural Characterization

### 3.2.1. XRD on the cell

For the cathode electrodes, the structure is Rhombohedral (R-3m space group), with the following lattice parameters:

- Pristine NMC cathode: a = b = 2.86461 Å, c = 14.27308 Å,
- Dead NMC cathode: a = b = 2.86563 Å, c = 14.28175 Å,

For the anode electrodes, the structure is hcp -hexagonal close packed (194:P63/mmc space group) with the following lattice parameters:

- Pristine graphite anode: a = b = 2.47758 Å, c = 6.72064 Å
- Dead graphite anode: a = b = 2.477946 Å, c = 6.72379 Å

All data were collected in the discharged state. The samples include a pristine electrode (after one cycle) and a degraded/dead electrode (after extended cycling at high C-rates until the capacity dropped below 42 mAh).

Fig. 3 (a) illustrates that the XRD plot for the NMC pristine and dead cathode exhibits a slight shift of the (003) peak to a lower 2θ in the dead cathode, indicating structural changes upon lithium intercalation/deintercalation. This shift suggests the presence of a reversible phase transformation and a reduction in the density of the dead cathode, associated with volume expansion, as further confirmed by EDS results in Section 3.2.3. [31, 32]

The increase in the c-lattice parameter (from 14.27308 Å to 14.28175 Å) indicates lithium removal from the interlayer spacing, which increases oxygen-oxygen repulsion between neighboring transition metal layers and results in expansion along the c-axis [33].

In Fig. 3 (b), the XRD analysis reveals notable structural changes in the anode. The enhanced peak intensity at 43.3° in the dead graphite anode corresponds to $Li_2CO_3$ surface species, as confirmed by XPS analysis (Fig. S1), indicating accumulation of surface carbonate species compared to the pristine graphite anode. At a 2θ value of 74.3°, a possible phase change is observed, as evidenced by the splitting of the peak. An increase in the c lattice parameter of graphite after extended cycling (from 6.72064 Å to 6.72379 Å) suggests interlayer expansion, likely due to lithium trapping, SEI thickening that reduces the total available Li and limits intercalation, or structural disorder induced by prolonged electrochemical stress.

### 3.2.2. SEM on electrodes

Fig. 4 (a) & (b) show top-view SEM images of the pristine and cycled NMC cathode electrodes, respectively. Initially, the cathode particles appear uniform in size, with clear grain boundaries and dense packing. In the pristine sample (Fig. 4a), the surface appears smooth, whereas the post-cycling sample (Fig. 4b) exhibits surface decomposition products and micro-cavities. These features suggest oxygen release, which is supported by the EDS composition (Table 2) showing a 2.7 at.% increase in oxygen content, indicative of structural degradation from prolonged cycling.

The pristine graphite electrode (Fig. 4c) displays a well-defined, layered morphology with smooth surfaces and minimal structural defects. After extended cycling (Fig. 4d), the graphite surface appears rougher, with visible cracks and edge exfoliation, indicating mechanical stress and the formation of the SEI layer.

### 3.2.3. EDS on electrodes

The EDS deconvolution spectrum (Fig. 5a) and elemental composition analysis (Table 2) show that the pristine cathode contains Ni (~19.53 wt.%), Mn (~6.67 wt.%), Co (~4.77 wt.%), and O (~22.18 wt.%), C (~10.05 wt.%), F (~6.46 wt.%), P (~0.24 wt.%). In contrast, the degraded cathode exhibits lower wt. % of these elements (Fig. 5b), indicating the formation of surface compounds such as $Li_2CO_3$ and amorphous carbon.

The density used in this work was obtained from the Bruker EDS software. This value is calculated based on the quantified elemental composition and X-ray absorption, assuming that the density contribution of each element corresponds to the tabulated density of the pure element. Accordingly, the total density is estimated as the weighted sum of the individual elemental densities (e.g., A × wt.% + B × wt.% + C × wt.%). To verify this value, we performed an independent calculation in Schulz et al. paper [15]. In this paper, the modeled depth was 800 nm and an energy collection up to 500 keV for the positive electrode, with a resonance energy of 441 keV. Subtracting these values yields an effective energy loss of ~59 keV. Using a stopping power (dE/dx) of 7.38 eV $Å^{-1}$ and SRIM software, the calculated density was 2.225 g $cm^{-3}$, which is in good agreement with the density estimated by the Bruker EDS software. The density information is provided in Figure S5.

The calculated density obtained from EDS analysis decreases after extended cycling from $4.624 \pm 0.07$ g/$cm^3$ to $4.20 \pm 0.02$ g/$cm^3$. This reduction can be attributed to the decreased weight percentage of heavy transition metals (Ni, Co, Mn) (Table 2), mainly due to the masking effect of the CEI layer. Tables S1 and S2 summarize the elemental and chemical state quantification derived from the fitted XPS spectra of Li, C, O, F, and P for the pristine and degraded anodes, respectively. The observed differences in elemental composition and chemical states indicate the formation and thickening of a carbonaceous surface layer during prolonged cycling, leading to an increased relative carbon content and attenuation of X-ray emission from subsurface species, as confirmed by cross-sectional analysis (Table S4). This shift in surface composition is consistent with surface reconstruction and phase transformations commonly observed in NMC cathodes under prolonged high-voltage cycling, further contributing to structural degradation.

The pristine anode (Table 3 & Fig. 6a) primarily contains C (~68.51 wt.%) with O (~13.80 wt.%) and F (~16.15 wt.%), P (~1.53 wt.%), whereas the cycled anode (Fig. 6b) shows an increased carbon signal with corresponding decreases in the weight percentages of oxygen, fluorine, and phosphorus. Since EDS reports relative elemental composition, an increase in surface carbon lowers the measured wt. % of other elements, even if their absolute amounts remain unchanged. These results suggest surface layer growth particularly an increase in the $Li_2Co_3$ peak intensity in the C 1s spectrum of the dead anode, rather than bulk loss of O, F, or P (Fig. S1).

The higher calculated density of the extended-cycle anode from $1.99 \pm 0.04$ g/cm$^3$ to $2.03 \pm 0.11$ g/cm$^3$, as compared to the one-cycle anode, can be attributed to lithium accumulation during prolonged cycling, as confirmed by Li-NRA measurements (Fig. 7d). Because lithium is a light element and is not detectable by EDS, its presence does not appear in the compositional spectra. Nevertheless, the buildup of lithium within the SEI and near-surface regions can increase the overall mass per unit volume, thereby contributing to the observed density increase. In addition, morphological changes such as SEI densification and reduced porosity may further enhance the measured density.

## 3.3. Depth Profile of Lithium by Li-NRA

To understand the bulk lithium distribution in commercial lithium-ion cells, Li-NRA was conducted on the 3032 coin cell using proton-induced gamma-ray emission (PIGE) to detect the $^{7}$Li concentration in both electrodes. Measurements were performed on electrodes under two conditions: a pristine electrode and a degraded electrode in the discharged state for safety. Raw data - ROI integral (counts) versus proton energy (MeV) were collected for the cathode (Fig. 7a) and anode (Fig. 7b), and used to calculate lithium content versus depth, as shown in Fig. 7c and 7d.

NRA analysis revealed changes in average lithium concentration after extended cycling for cells with an NMC cathode paired with a graphite anode: the NMC cathode exhibited a bulk lithium loss of 1.86 x $10^{22}$ atoms/$cm^3$ while the graphite anode showed a corresponding gain of 1.38 x $10^{22}$ atoms/$cm^3$ (Fig. 7c and 7d), indicating irreversible lithium consumption and structural degradation over extended cycling.

In the degraded anode, an increase in density - observed by EDS spectroscopy and averaged over five different surface points that correlates with reduced lithium penetration depth compared to the pristine anode (Fig. 7d). Conversely, in the dead cathode, a decrease in density - also determined by EDS and averaged over five surface points, which corresponds to deeper lithium penetration in the Li-NRA depth profile relative to the pristine cathode (Fig. 7c).

These results demonstrate the capability of Li-NRA to provide spatially resolved lithium quantification and highlight its value in diagnosing degradation pathways in commercial battery systems.

# 4. Conclusions

In summary, high C-rate cycling substantially accelerates the degradation of Ni-rich NMC cathodes. Our electrochemical measurements show that 3C charging leads to a significantly faster capacity loss (~ 60.19%) and higher internal resistance (0.34 Ω) compared to 1C & 2C rates. Structural analysis revealed an expansion of the c-axis lattice parameter in the degraded cathode, consistent with lithium loss measured by Li-NRA (~19.7%).

On the anode side, lithium plating on the graphite electrode is the main aging mechanism, even at room temperature. Although capacity fade per cycle may seem smaller at higher C-rates, Li-NRA shows that Li migration causes a significant loss of cyclable lithium, creating a capacity imbalance between electrodes. High C-rate cycling, combined with large IR drop and limited $Li^+$ transport, accelerates anode degradation, leading to capacity fade, increased internal resistance (from SEI growth and loss of active Li), and lower coulombic efficiency. Quantitatively, the degraded graphite anode shows a 115.96% increase in lithium content, reflecting severe lithium trapping and plating.

The non-destructive Li-NRA method is an effective technique for probing the lithium depth profile, particularly given lithium's high mobility. It enables quantitative determination of lithium content in both charged and discharged states, which can be directly compared with XRD measurements to track changes in the c-lattice parameter of the cathode as lithium is extracted during charging and reinserted during discharging. A future direction of this work is to collect Li-NRA data separately for charged and discharged states at different C-rates, providing a comprehensive picture of lithium distribution under high C-rate cycling.

# 5. Acknowledgements

This research project was made possible by the generous support of the NYS Center for Advanced Technology in Nanomaterials and Nanoelectronics (CATN2). We are deeply grateful for their financial contribution and unwavering belief in our work, which has been instrumental in achieving our research goals. I am also thankful to Valenti Anthony, Robert Bohl, Dr. Wei Zheng, and Dr. Sandra Schujman for their assistance with data collection.

# 6. Declaration of generative AI and AI-assisted technologies in the writing process

During the preparation of this work, the author(s) used ChatGPT in order to find and fix grammatical errors, voice dictation, and improve the readability of the manuscript. After using this tool/service, the author(s) reviewed and edited the content as needed and take(s) full responsibility for the content of the published article.

# Figures

## Figure 1

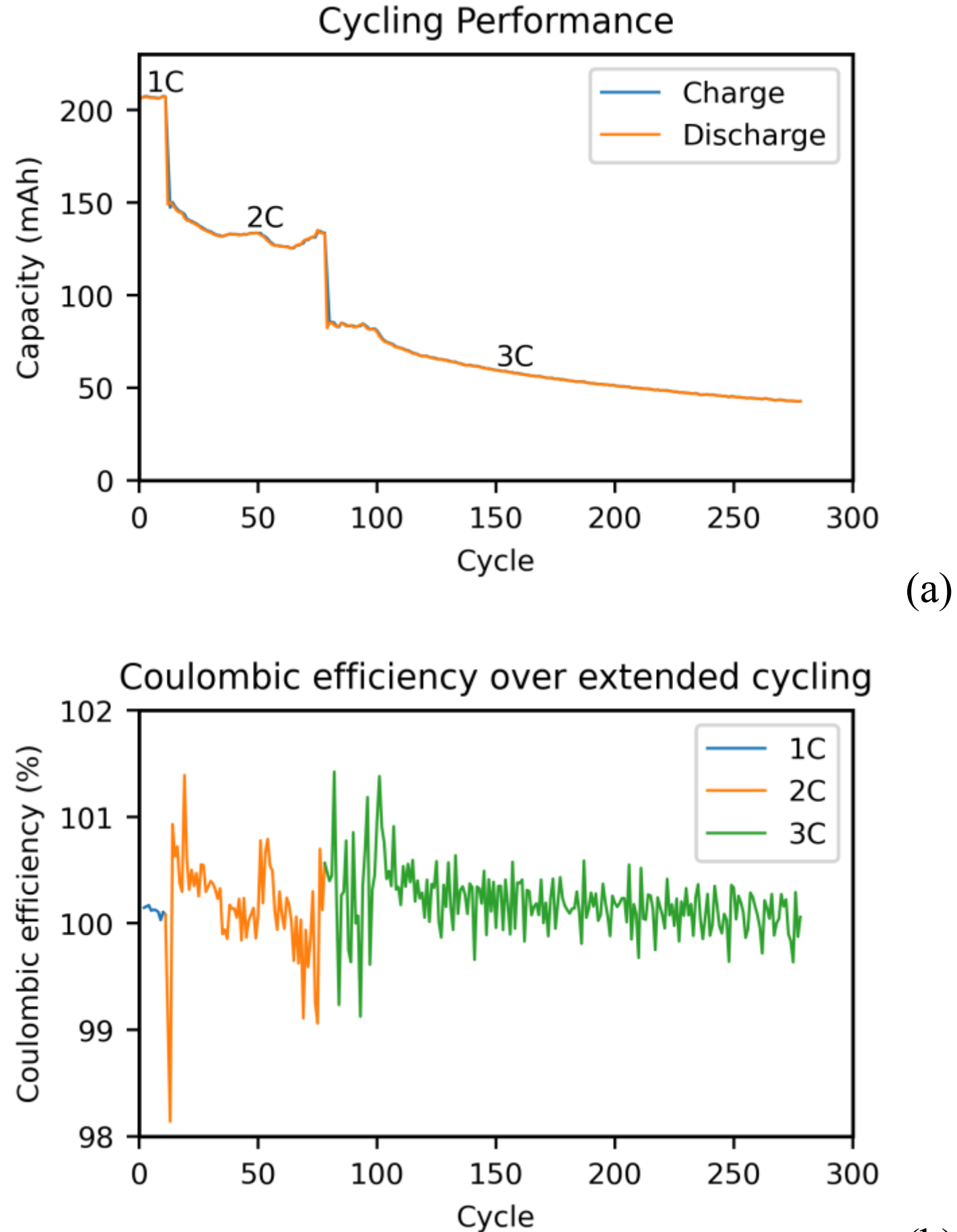

**Figure 1: Cycling Performance and Coulombic efficiency** : (a) Capacity versus cycle number for Graphite/NMC commercial coin cell. Cycling was conducted between 2.7 and 4.3 V at 25 °C with 1C (200 mA), 2C (400mA), 3C (600mA) (b) C-rate dependent Coulombic Efficie

Figure 2

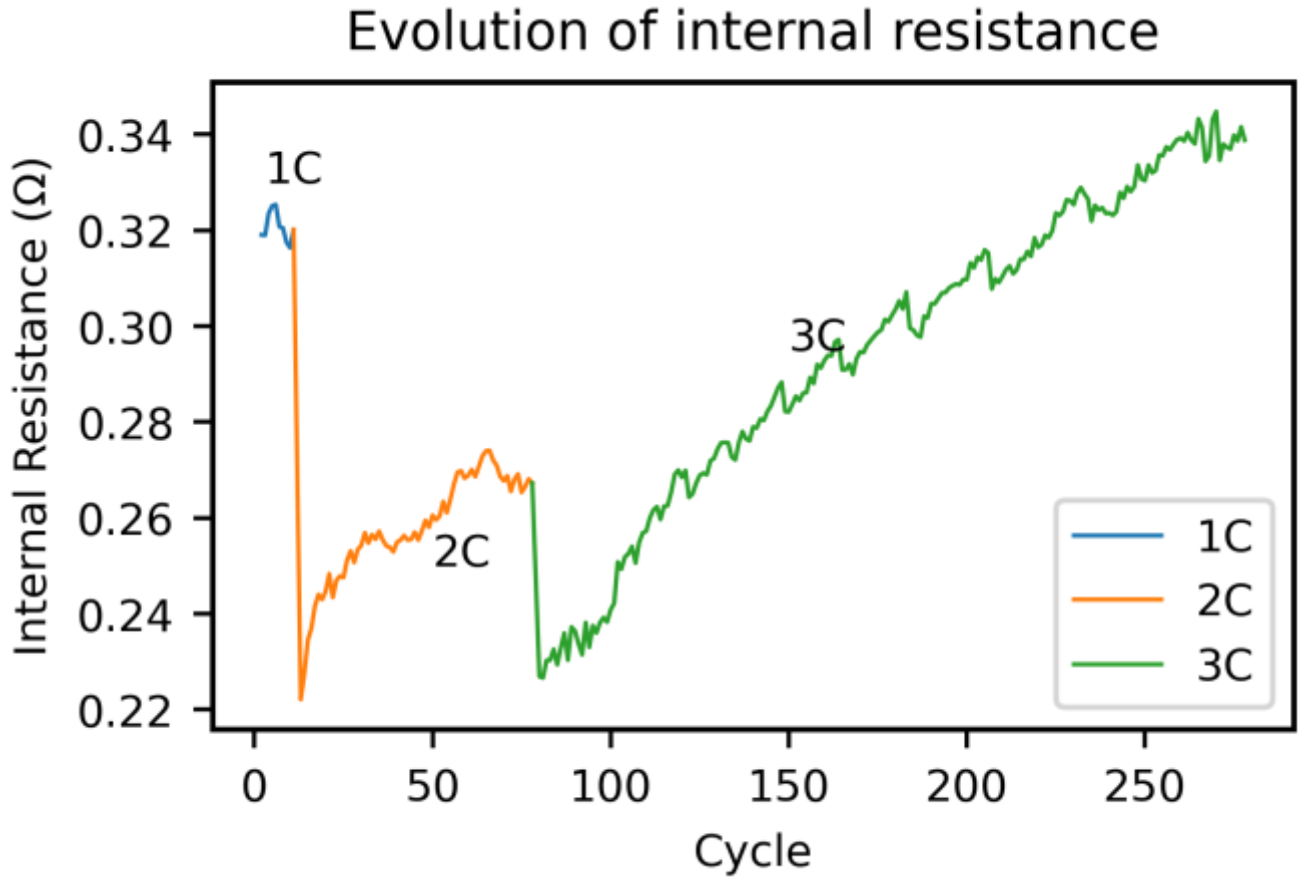

**Figure 2: Evolution of internal resistance:** NMC/Graphite coin cell across 1C, 2C, and 3C cycling rates, showing initial resistance drops upon rate changes and a progressive linear increase at 3C, correlating with capacity fade.

# Figure 3

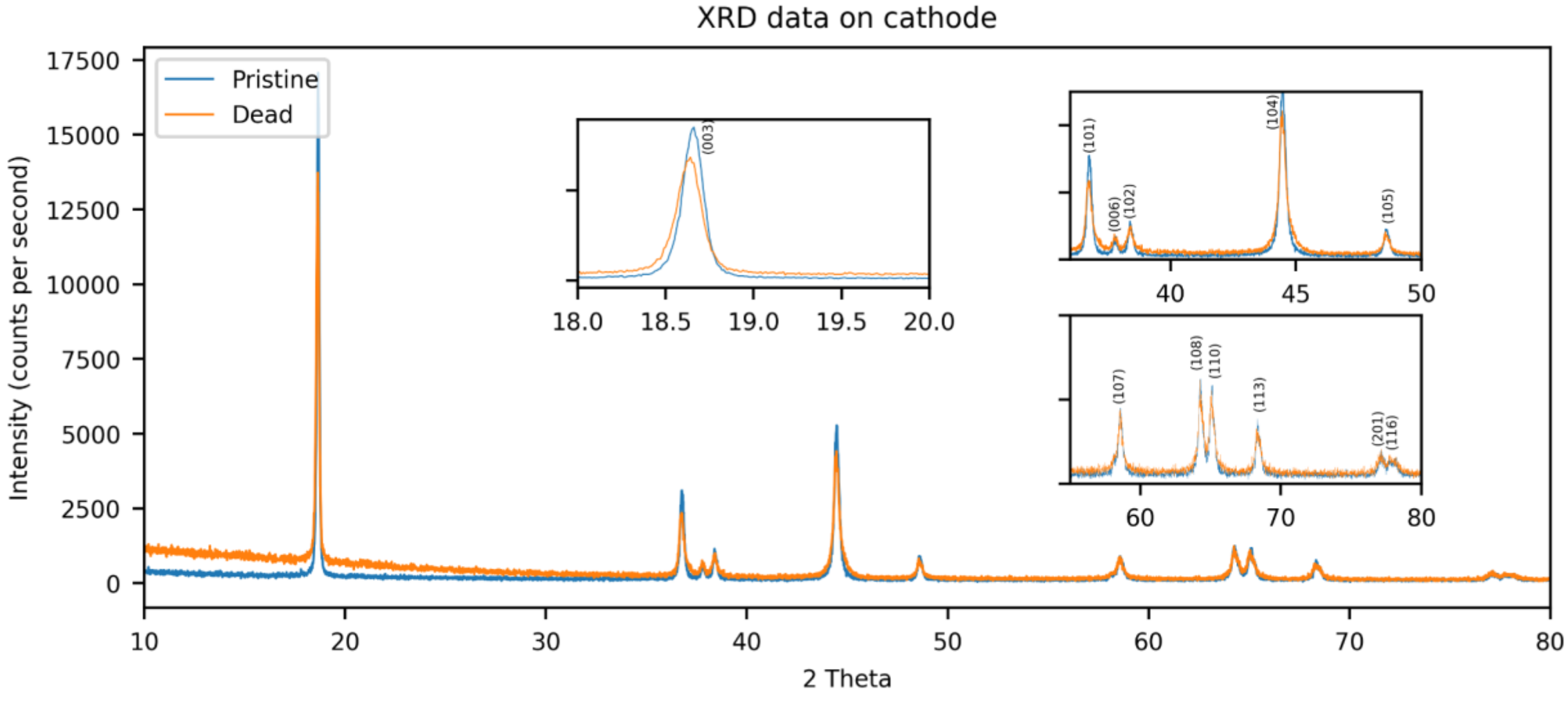

(a)

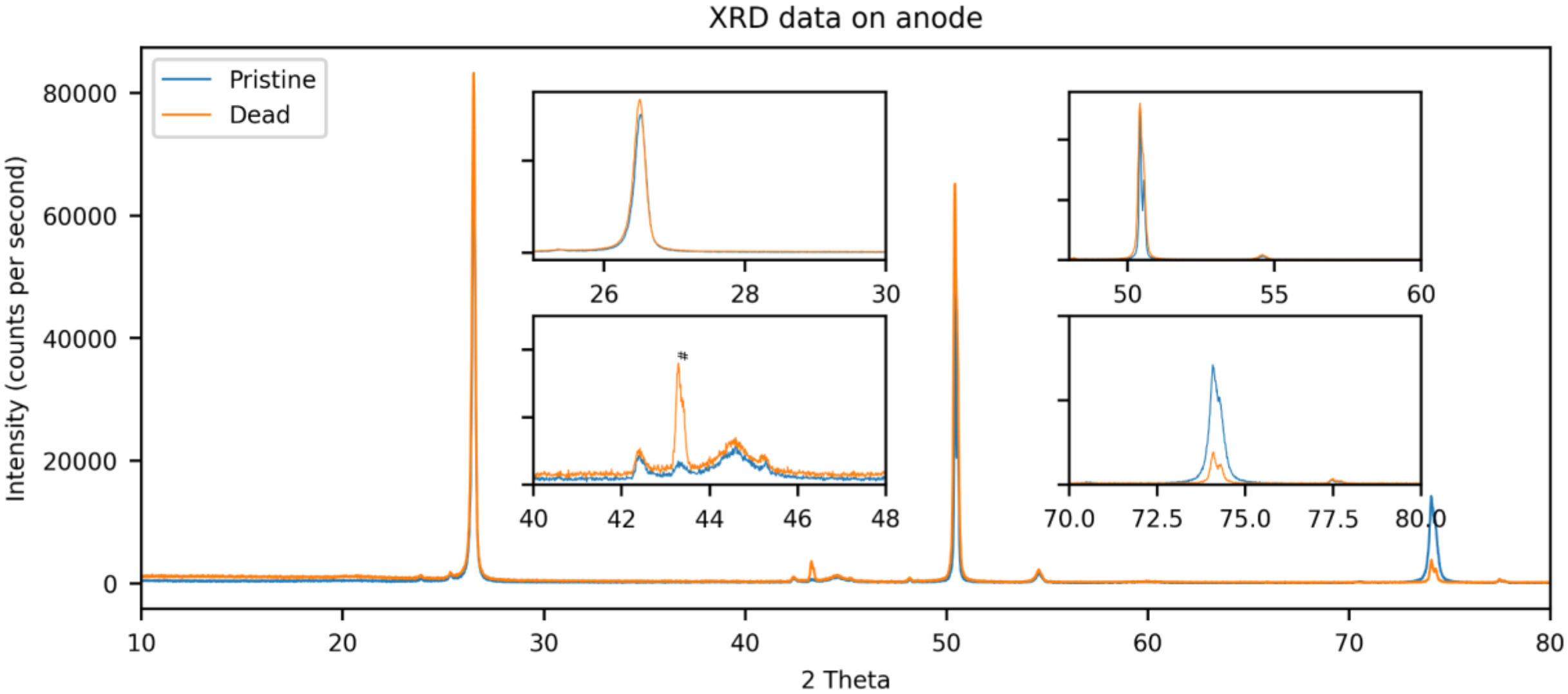

(b)

**Figure 3: XRD data:** on (a) cathode: Pristine and extended cycled (dead) cathode (NMC) with insight diagram (003) peak shifted slightly towards a lower angle, and (b) anode: Graphite peak enhanced at 43.3° & peak splitting observed at 74.3° after 278 cycles.

Figure 4

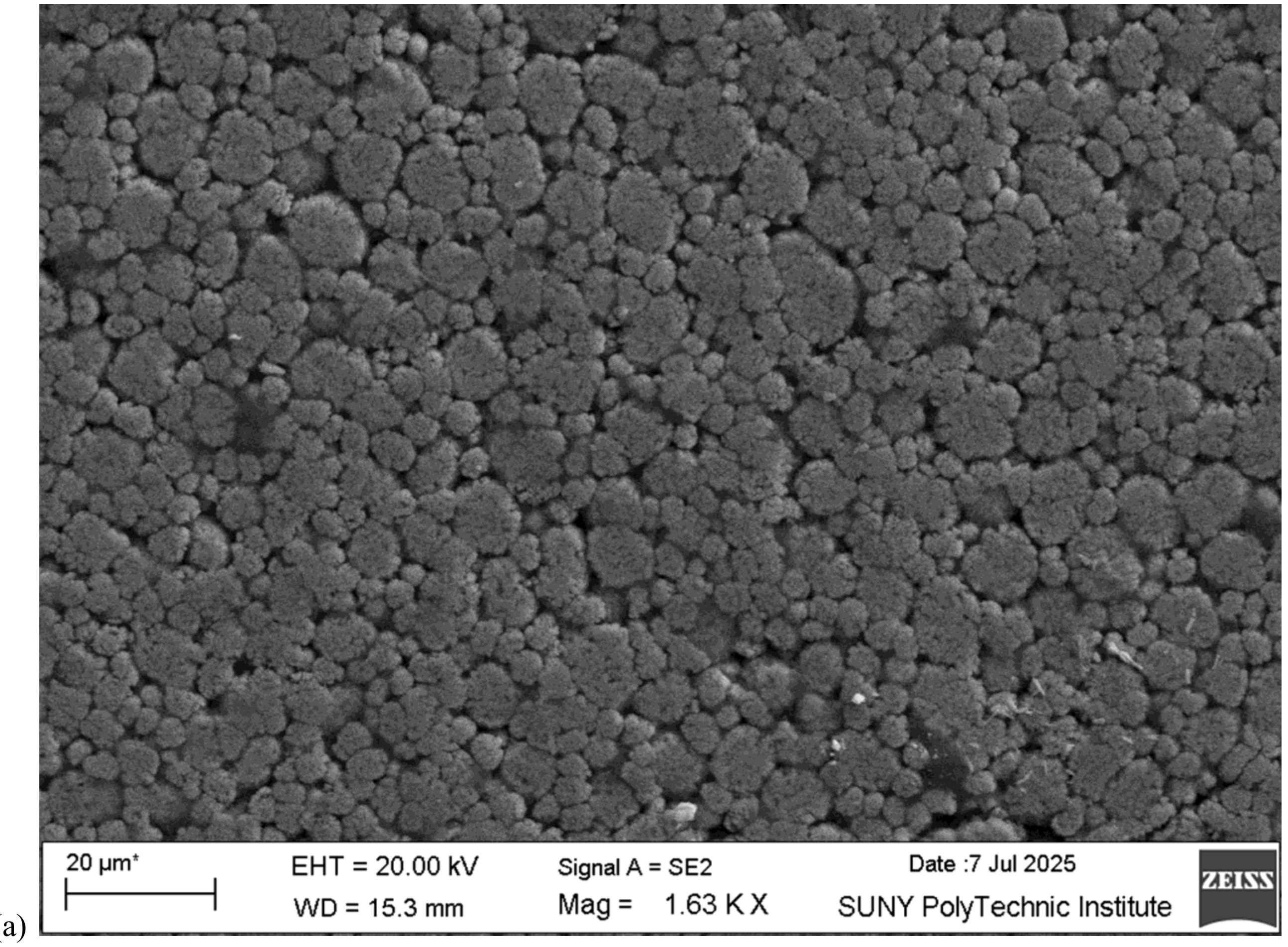

(a)

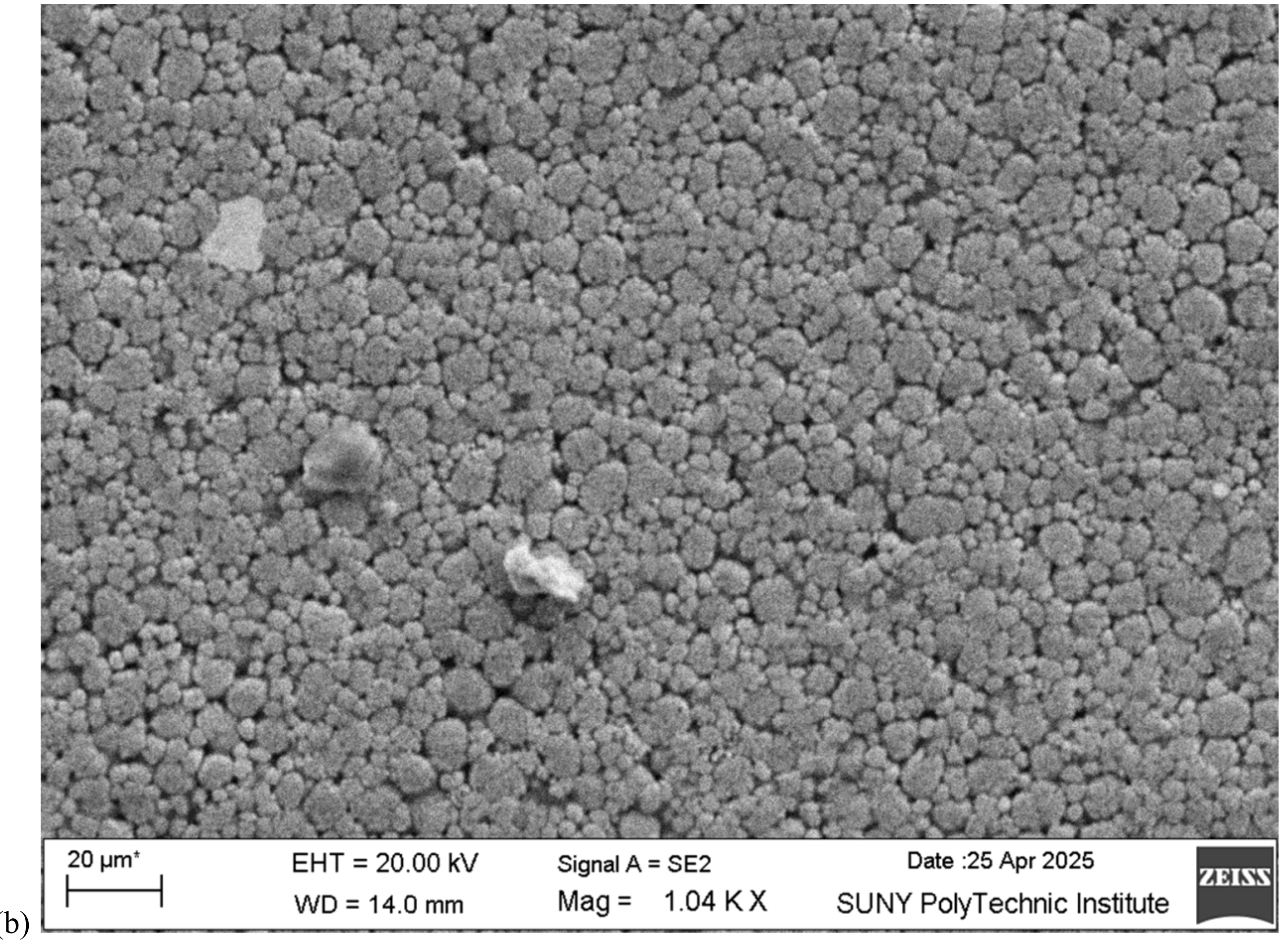

(b)

(c) 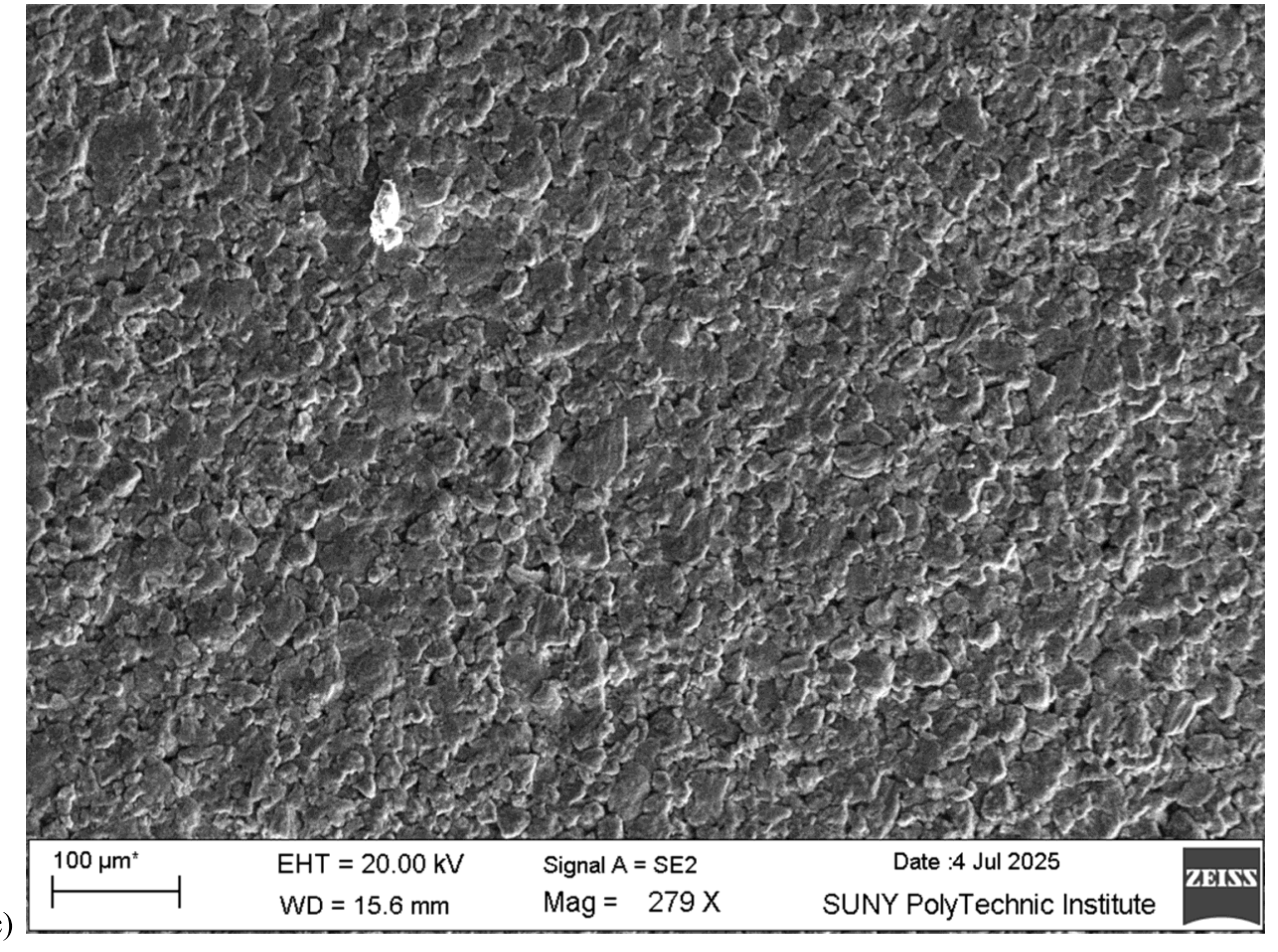

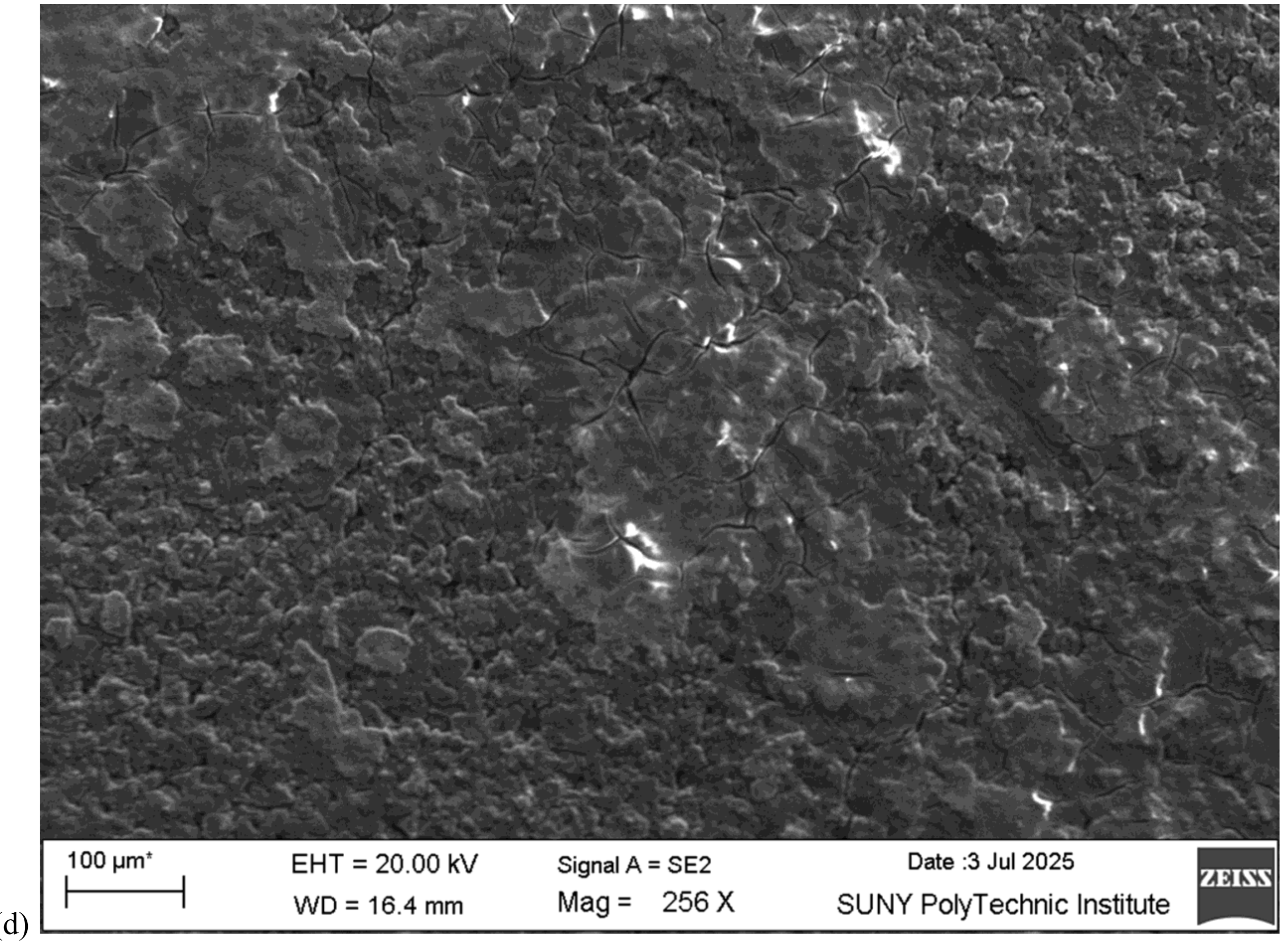

**Figure 4: SEM images of NMC/Graphite electrodes:** a&b) 1000x magnification (a) pristine and (b) cycled NMC cathodes; c&d) 100x magnification (c) pristine and (d) cycled graphite anodes. Cycling induces surface degradation and micro-cavities in the cathode, and roughening, cracking, and exfoliation in the anode.

Figure 5

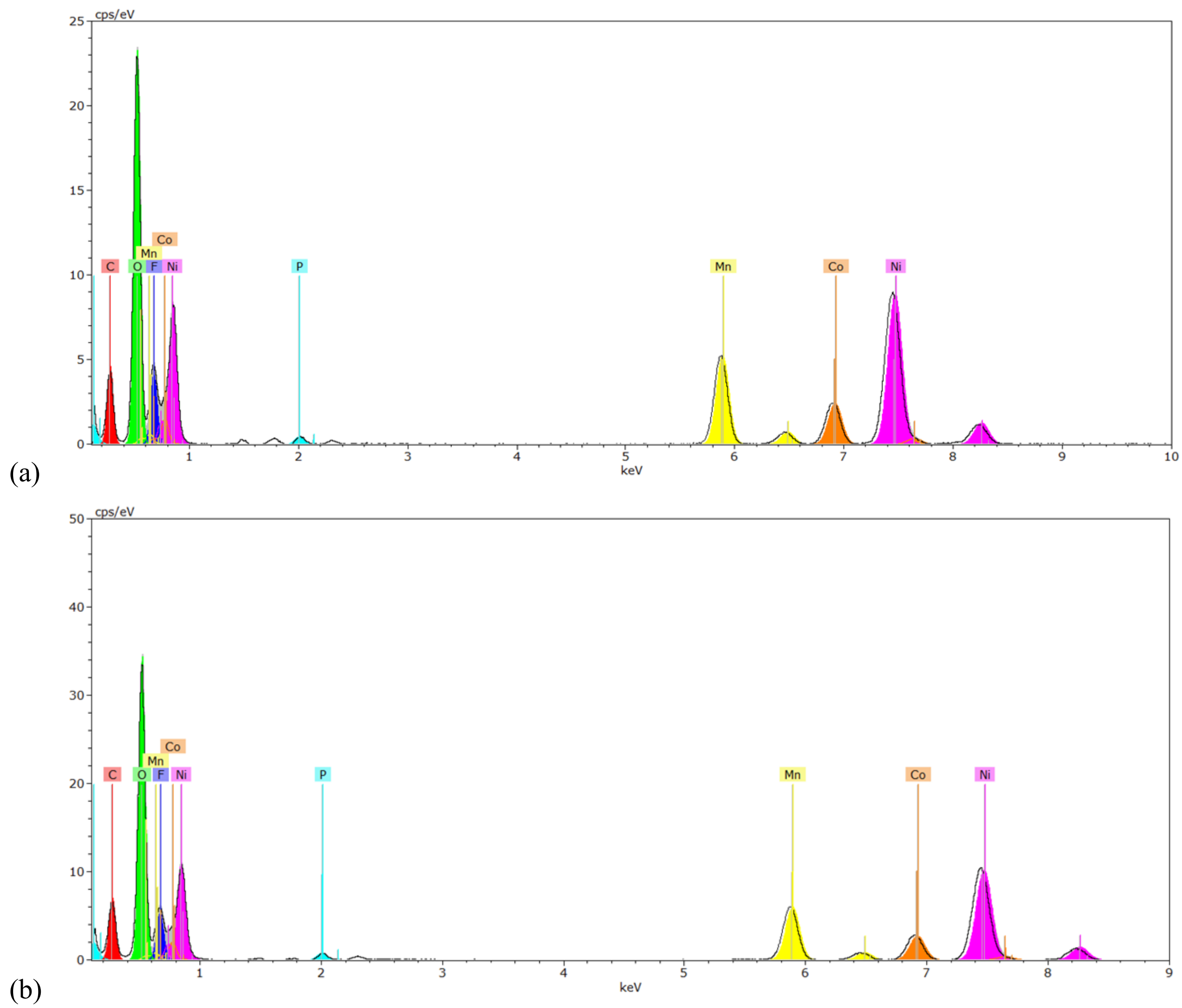

**Figure 5: EDS deconvolution spectra of cathode:** (a) pristine NMC and (b) degraded NMC.

Figure 6

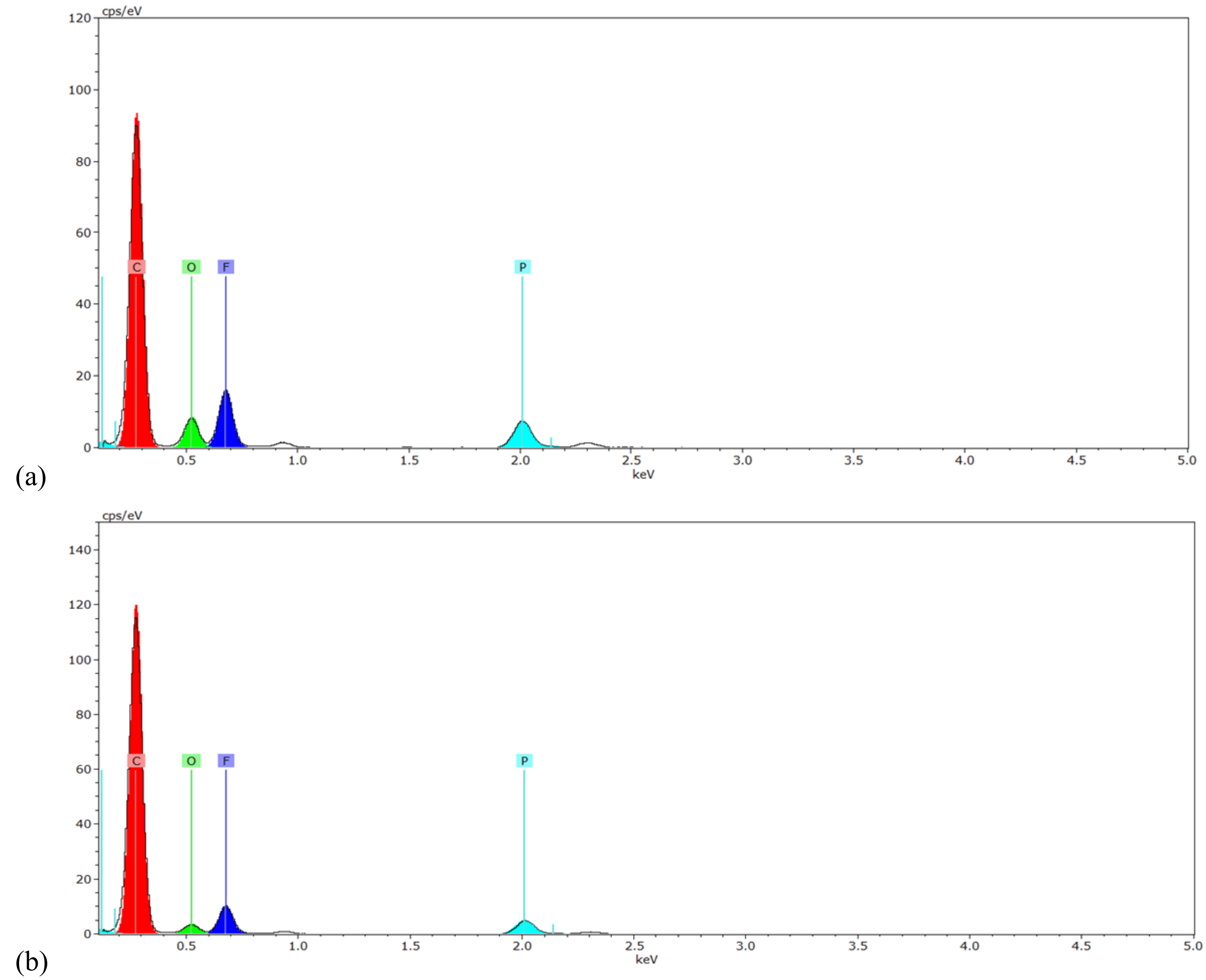

**Figure 6: EDS deconvolution spectra of anode:** (a) pristine graphite and (b) cycled graphite

Figure 7

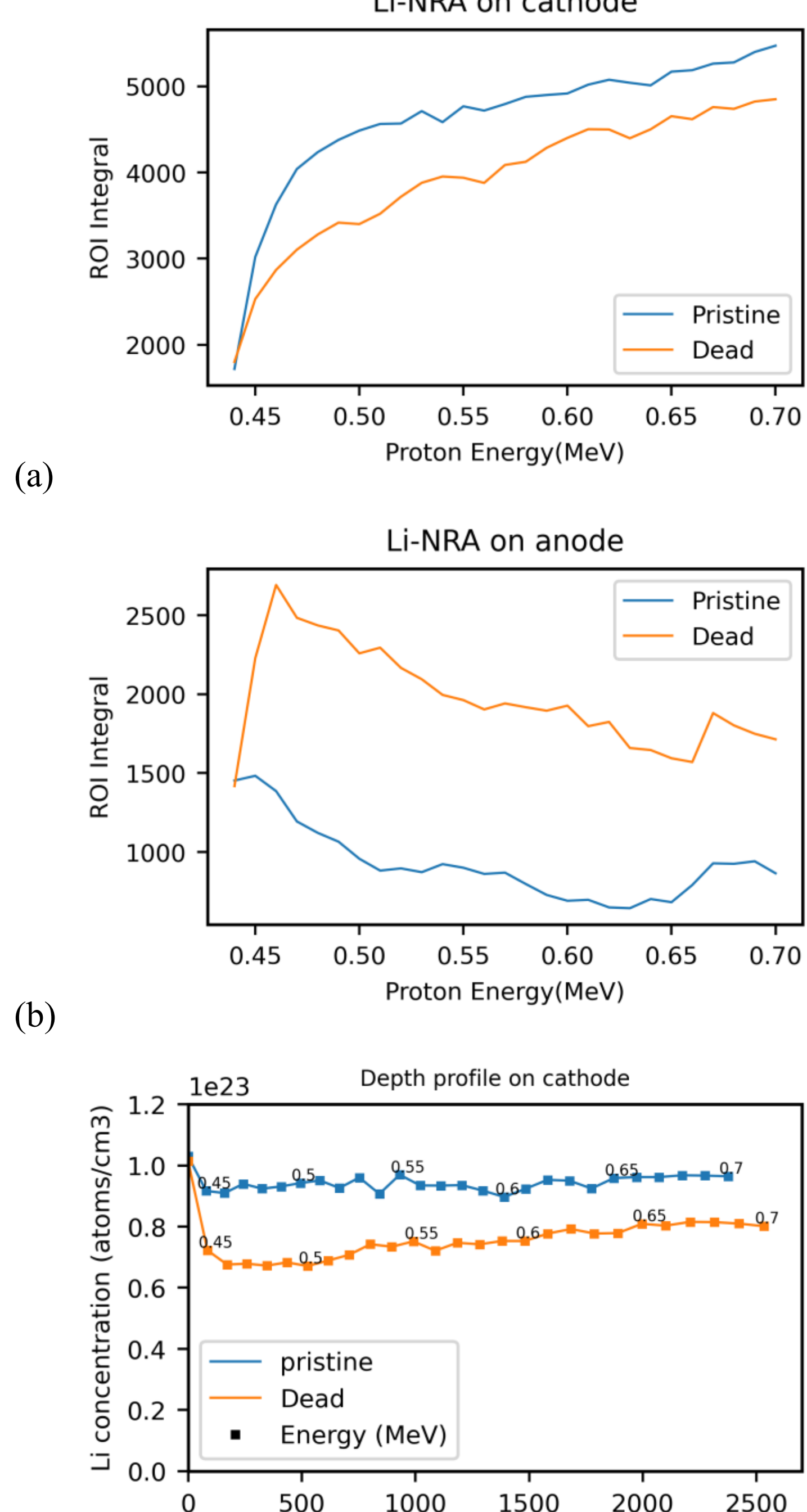

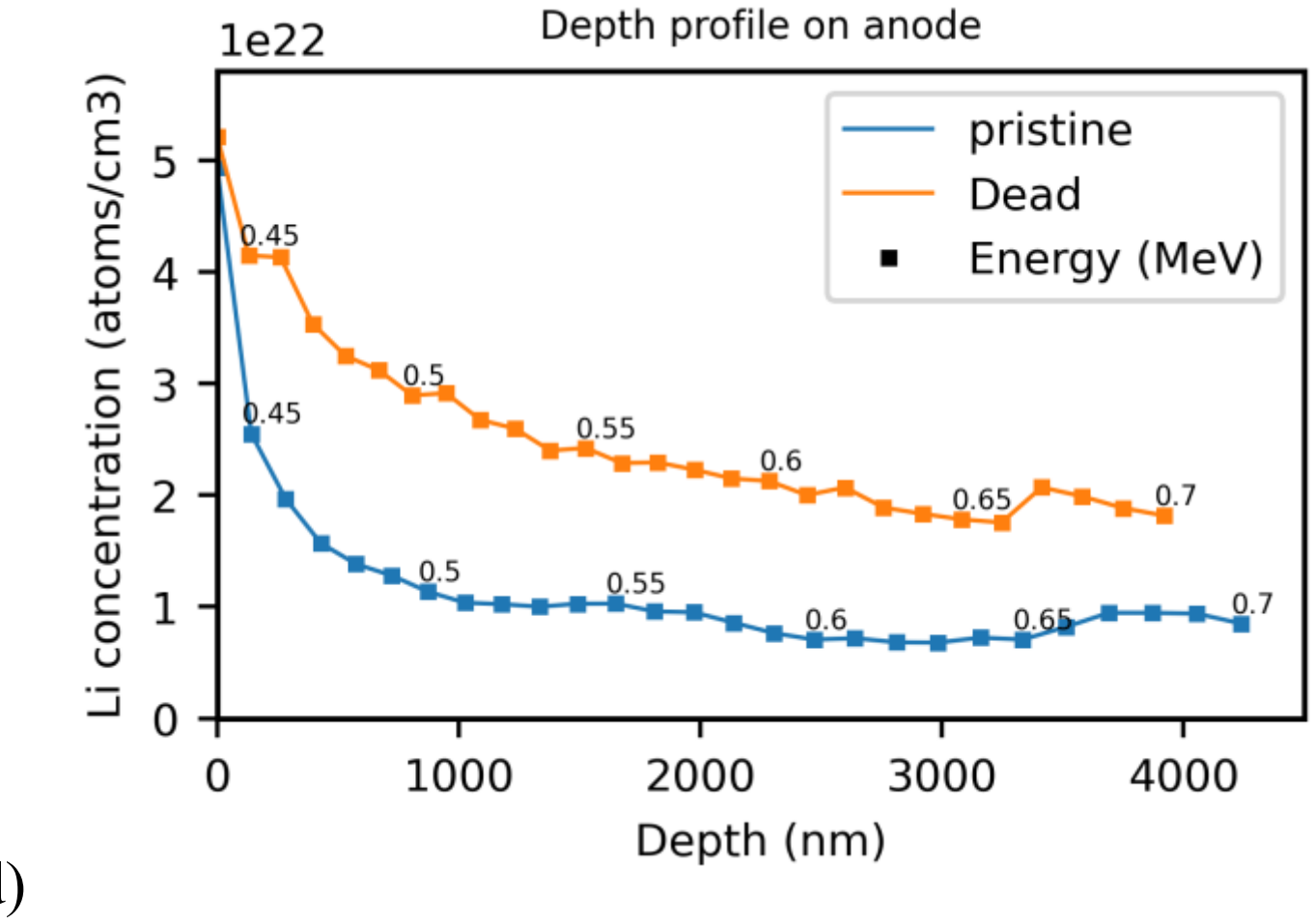

(d)

**Figure 7: Li-NRA and depth profile on electrodes:** (a) and (b) show the variation in gamma-ray intensity as a function of incident proton energy (increased in 10 keV steps) for pristine and dead cathode and anode, respectively. (c) and (d) present the corresponding lithium depth profiles obtained from NRA analysis. The numbers indicated within the depth profile diagrams represent the proton energy (in MeV) associated with that specific depth.

# Tables

## Table 1

| C-rate | Cycles (Used for calc.) | Start capacity (mAh) | End capacity (mAh) | Total Retention (%) | Fade (mAh/cycle) | Fade (%/cycle) |
|---|---|---|---|---|---|---|
| 1-C | 11 | 206.5 | 207.06 | 100.27 | ~0.05 | ~0.002 |
| 2-C | 67 | 148.97 | 132.84 | 89.17 | 0.241 | 0.162% |
| 3-C | 200 | 82.16 | 42.67 | 51.94 | 0.197 | 0.240% |

**Table 1: Capacity retention and fade rate:** NMC/graphite coin cell at different C-rates. Retention values were calculated by normalizing the discharge capacity at the end of each C-rate cycling block to the initial discharge capacity within that block. Both absolute fade rates (mAh per cycle) and normalized fade rates (% per cycle) are reported, demonstrating that higher C-rates lead to increased relative capacity loss despite sometimes lower absolute capacity fade per cycle.

Table 2

| Element (Series) | unn.C [wt.%] | | norm.C [wt.%] | | Atom.C [at.%] | | Error($3\sigma$) [wt.%] | |
|---|---|---|---|---|---|---|---|---|
| | pristine | degraded | pristine | degraded | pristine | degraded | pristine | degraded |
| Carbon-K | 10.05 | 7.35 | 14.38 | 15.95 | 26.94 | 27.98 | 3.60 | 2.57 |
| Oxygen-K | 22.18 | 16.56 | 31.73 | 35.94 | 44.63 | 47.33 | 7.21 | 5.33 |
| Fluorine-K | 6.46 | 4.31 | 9.24 | 9.36 | 10.95 | 10.38 | 2.34 | 1.56 |
| Phosphorus-K | 0.24 | 0.25 | 0.34 | 0.55 | 0.25 | 0.37 | 0.10 | 0.11 |
| Nickel-K | 19.53 | 10.59 | 27.94 | 22.99 | 10.71 | 8.25 | 1.63 | 0.92 |
| Manganese-K | 6.67 | 4.34 | 9.54 | 9.43 | 3.91 | 3.61 | 0.61 | 0.42 |
| Cobalt-K | 4.77 | 2.66 | 6.83 | 5.78 | 2.61 | 2.07 | 0.46 | 0.29 |
| Total | 69.90 | 46.08 | 100.00 | 100.00 | 100.00 | 100.00 | | |

**Table 2: Elemental composition:** Pristine/Degraded NMC cathode by EDS

Table 3

| Element (Series) | unn.C [wt.%] | | norm.C [wt.%] | | Atom.C [at.%] | | Error(3$\sigma$) [wt.%] | |
|---|---|---|---|---|---|---|---|---|
| | pristine | degraded | pristine | degraded | pristine | degraded | pristine | degraded |
| Carbon-K | 68.51 | 80.55 | 68.51 | 80.55 | 76.40 | 86.18 | 21.28 | 24.89 |
| Oxygen-K | 13.80 | 7.35 | 13.80 | 7.35 | 11.55 | 5.90 | 4.63 | 2.61 |
| Fluorine-K | 16.15 | 11.06 | 16.15 | 11.06 | 11.39 | 7.48 | 5.27 | 3.68 |
| Phosphorus-K | 1.53 | 1.05 | 1.53 | 1.05 | 0.66 | 0.44 | 0.25 | 0.20 |
| Total | 100.00 | 100.00 | 100.00 | 100.00 | 100.00 | 100.00 | | |

**Table 3: Elemental composition:** Pristine/Degraded graphite anode by EDS